\documentclass[a4paper,10pt]{IEEEtran}
\usepackage[left=1.3cm,right=1.3cm,top=1.82cm, bottom = 4.3cm]{geometry}

\usepackage{setspace}
\usepackage{multirow}
\usepackage{lipsum}
\usepackage{amsfonts}
\usepackage{array}
\usepackage{amsmath,cases,amssymb}
\usepackage{amsmath}
\usepackage{amsthm}
\usepackage{graphicx}
\usepackage{cite}
\usepackage{subfigure}
\usepackage{epsfig}
\usepackage[hyphens]{url}
\usepackage{algorithm}
\usepackage{algorithmic}
\usepackage{epstopdf}
\usepackage{color}
\usepackage{makecell}
\usepackage{xcolor}
\usepackage[normalem]{ulem}

\usepackage{mathtools}

\DeclareGraphicsExtensions{.eps,.pdf}

\usepackage{soul}

\makeatletter
\newcommand{\doublewidetilde}[1]{{%
		\mathpalette\double@widetilde{#1}}}
\newcommand{\double@widetilde}[2]{%
	\sbox\z@{$\m@th#1\widetilde{#2}$}%
	\ht\z@=.5\ht\z@
	\widetilde{\box\z@}}
\makeatother

\usepackage{glossaries}
\usepackage{tikz}
\usetikzlibrary{shapes.arrows}
\usetikzlibrary{plotmarks}
\usetikzlibrary{patterns}
\usetikzlibrary{arrows.meta}
\usetikzlibrary{positioning}
\usetikzlibrary{spy}
\usepackage{pgfplots}
\usepgfplotslibrary{colorbrewer}
\usepgfplotslibrary{fillbetween}
\usepackage{subcaption} 
\usepackage{balance}
\usepackage{amsmath}

\usepackage{soul}

\usepackage{pdfpages}
\begin{document}
\newpage
\newacronym{RAN}{RAN}{Radio Access Network}
\newacronym{ORAN}{ORAN}{Open Radio Access Network}
\newacronym{GAI}{GAI}{Generative AI}
\newacronym{MEC}{MEC}{Mobile Edge computing}
\newacronym{UE}{UE}{User Equipment}
\newacronym{UL}{UL}{Uplink}
\newacronym{DL}{DL}{Downlink}
\newacronym{MCS}{MCS}{Modulation and coding scheme}
\newacronym{RIC}{RIC}{RAN Intelligent Controller}

\newacronym{CN}{CN}{Core Network}

\newacronym{nonrt}{non-RT}{non-real-time}
\newacronym{nrt}{near-RT}{near-real-time}

\newacronym{capex}{CapEx}{Capital Expenditure}
\newacronym{opex}{OpEx}{Operational Expenditure}
\newacronym{ML}{ML}{Machine Learning}
\newacronym{SM}{SM}{Service Model}
\newacronym{UAV}{UAV}{Unmanned Aerial Vehicle}

\newacronym{URLLC}{URLLC}{Ultra Reliable Low-Latency Communication}
\newacronym{EMBB}{eMBB}{Enhanced Mobile BroadBand}
\newacronym{PSNR}{PSNR}{Peak signal-to-noise ratio}
\newacronym{VMAF}{VMAF}{Video Multi-Method Assessment Fusion}
\newacronym{MSE}{MSE}{Mean Squared Error}
\newacronym{SLA}{SLA}{Service Level Agreement}
\newacronym{QOE}{QoE}{Quality of Experience}
\newacronym{SNR}{SNR}{Signal-to-Noise-Ratio}
\newacronym{KPI}{KPI}{Key Performance Indicator}
\newacronym{ABR}{ABR}{Adaptive Bit Rate}
\newacronym{RTI}{RTI}{Real-Time Interaction}

\newacronym{BS}{BS}{Base Station}
\newacronym{PRB}{PRB}{Physical Resource Block}
\newacronym{GPU}{GPU}{Graphical Processing Unit}
\newacronym{SSIM}{SSIM}{structural similarity}

\newacronym{iot}{IoT}{Internet of Thing}
\newacronym{TX}{TX}{Transmitter}
\newacronym{RX}{RX}{Receiver}

\newacronym{mMIMO}{mMIMO}{massive MIMO}
\newacronym{NEF}{NEF}{Network Exposure Function}
\newacronym{PAA}{PAA}{Probe and Adapt Algorithm}
\newacronym{PCC}{PCC}{Proactive Congestion Control}
\newacronym{E2E}{E2E}{End-to-End}

\title{Experimental Study of Low-Latency Video Streaming in an ORAN Setup with Generative AI}

\author{Andreas Casparsen, Van-Phuc Bui, Shashi Raj Pandey, Jimmy Jessen Nielsen, Petar Popovski\\
Department of Electronic Systems, Aalborg University, Denmark, Emails: \{aca, vpb, srp, jjn, petarp\}@es.aau.dk
		\thanks{This work was supported by the Villum Investigator Grant ``WATER" from the Velux Foundation, Denmark}	}

\maketitle
\begin{abstract}
Current feedback-based congestion control methods, such as probe-and-adapt solution, for live video streaming react after it occurs, causing buffering and latency spikes. We introduce a proactive semantic control channel enabling coordination between \gls{ORAN} xApp, \gls{MEC}, and \gls{UE} for seamless mobile video streaming. When the transmitting UE experiences poor \gls{UL} conditions, the MEC proactively instructs to downscale video based on low-level RAN metrics—such as millisecond SNR updates, preventing buffering before it fully manifests.
A \gls{GAI} module at the \gls{MEC} reconstructs high-quality frames from downscaled video before forwarding them over the typically stronger \gls{DL}. Experiments on a live \gls{ORAN} testbed with 50 video streams show reduced latency tails and up to 4 dB PSNR and 15 \gls{VMAF} gains over reactive congestion control. The proactive control eliminates latency spikes over 600 ms, demonstrating effective cross-layer coordination for latency-critical streaming.

\end{abstract}
%%%%%%%%%%%%%%%%%%%%%%%%%%%%%%%%%%%%%%%%%%%%%%%
\section{Introduction}\label{sec:intro}
%%%%%%%%%%%%%%%%%%%%%%%%%%%%%%%%%%%%%%%%%%%%%%%
\glsresetall
The \gls{ORAN} paradigm \cite{10024837} offers a promising solution to address challenges of network resource management due to the growing demand for real-time data processing and communication across various Industry 4.0 scenarios through its flexible and software-defined architecture. Near-real-time (Near-RT) xApps, operating via the RIC, enable responsive RAN control at the 10-millisecond–1-second timescale, supporting procedures such as RAN slicing and load balancing. Their execution enhances traffic flows like EMBB and URLLC, improving overall bitrate and latency \cite{johnson2022nexran,wiebusch2023towards} of the \gls{RAN}.
Such dynamic resource-control capabilities offer new opportunities to improve multimedia transmission under challenging network conditions. 

Intelligent network management becomes particularly essential in scenarios where reliable, ultra-low-latency live video streaming with high visual fidelity is required despite highly variable link qualities. For example, live video streaming between mobile devices is driven by the growing demand for reliable, high-quality real-time multimedia services in situations with constrained network resources, such as industrial \gls{iot}, remote monitoring, and smart cities \cite{9877931}. Traditional solutions for live-video streaming include bitrate prediction and \gls{ABR} \cite{li2023fleet,taraghi2023lll} to reduce latency under fluctuating conditions, but depend on reactive solutions when the degradation in link quality is observed. Alternatively, \gls{PAA} can also be employed, which, through feedback-based congestion control measures, sets the available rate and compression rates \cite{li2014probe}.
These limitations are further exacerbated in \gls{iot} networks, where \gls{UL} transmission is more challenging than \gls{DL} transmission, due to the limited power budgets and hardware constraints of handheld \glspl{UE}.

\gls{GAI} can restore video quality after low-rate, lossy compression by applying suitable upscaling techniques, vital in constrained networks that cannot consistently support high-bitrate video streaming. Prior work has explored semantic and goal-oriented communication in \gls{ORAN} \cite{strinati2024goal} and applied \gls{GAI} to \gls{UAV} networks \cite{kaleem2024emerging}, but these solutions rely on non-real-time control and lack fine-grained adaptation.
Motivated by these limitations and \gls{ORAN}'s fine-grained control capabilities, we explore integrating \gls{ORAN}-based control with \gls{GAI} to enhance video streaming performance over constrained wireless links by leveraging semantic communication principles that reduce information redundancy and minimize latency. 
%\aca{
While our method reacts to real-time RAN conditions rather than predicting future states, it operates proactively as rate adaptation is triggered ahead of congestion effects visible at higher protocol layers. This enables semantic-driven behavioral adjustments for \gls{UE} video transmission based on real-time network feedback.%}
Our contributions are: 1) an ORAN-compliant architecture leveraging MEC and a semantic control channel for proactive video transmission adaptation based on RAN feedback, we refer to this as \gls{PCC}; 2) experimental validation on a live testbed \cite{githubrepo} demonstrating GAI-based quality improvements compared to typical \gls{PAA} solutions for feedback-based congestion control; and 3) analysis of latency-quality trade-offs in live video streaming scenarios.

The remainder of the paper is organized as follows: Section~\ref{sec:system} presents the system architecture, Section~\ref{sec:results} details the experimental setup, Section~\ref{sec:Numerical} provides performance evaluation, and Section~\ref{sec:conclusion} concludes.

\begin{figure}[t!]
	\centering
	\includegraphics[width=0.31\textwidth]{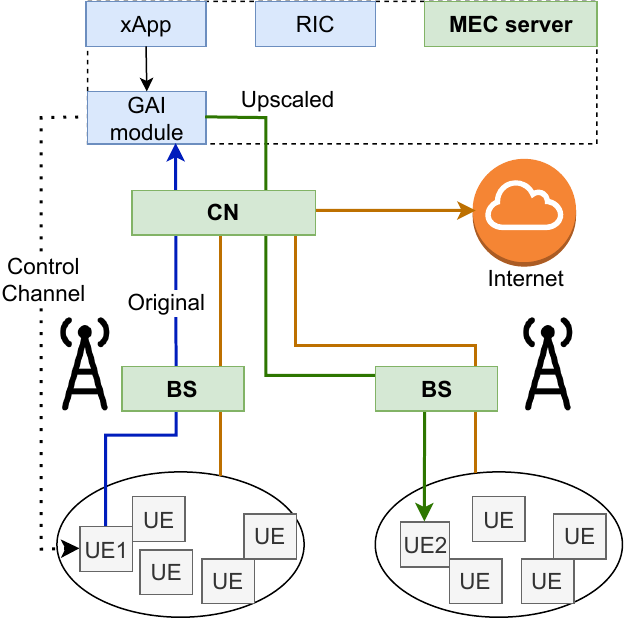}
    % \vspace*{-0.3cm}
 	\caption{Proposed system for GAI-based MEC and CN.}
	\label{fig:Systemmodel}
	\vspace{-15pt}
\end{figure}

%%%%%%%%%%%%%%%%%%%%%%%%%%%%%%%%%%%%%%%%%%%%%%%
\section{Intelligent RAN-Aware Architecture for Video Quality Optimization}\label{sec:system}
\begin{figure}[t!]
    \centering
        \includegraphics[width=0.85\linewidth]{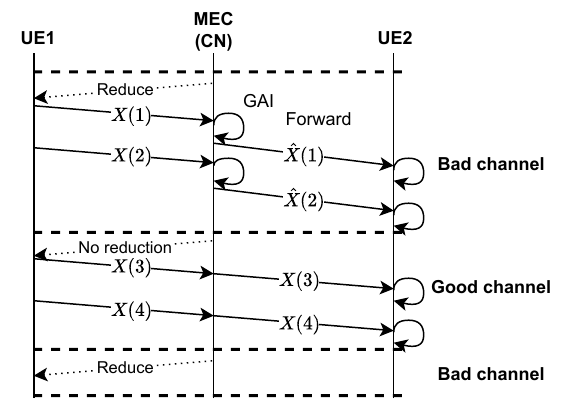}

    \caption{Application of GAI in a video streaming context.}
    % \vspace{-15pt}
    \label{fig:software}
    \vspace{-15pt}
\end{figure}

%%%%%%%%%%%%%%%%%%%%%%%%%%%%%%%%%%%%%%%%%%%%%%%
The proposed system architecture, illustrated in Fig.~\ref{fig:Systemmodel}, presents real-time video streaming between UEs possibly located in different cells. The key innovation is a \textit{semantic} control channel enabling tight coordination between the \gls{ORAN} xApp, \gls{MEC}, and UEs through the \gls{NEF}. Here \emph{semantic} reflects the fact that the MEC operation is content-aware and adjusts the operation accordingly. Due to power and hardware constraints of handheld UEs, \gls{UL} transmission is often more challenging than \gls{DL}, particularly when UE1 experiences degraded channel conditions. 
The xApp continuously monitors \gls{RAN} conditions and supplies real-time channel quality metrics updated on a millisecond timescale, supporting proactive \gls{MEC} adaptation. When channel conditions are insufficient, UE1 is instructed to compress and transmit video frames with reduced quality to avoid buffering and packet loss. This control channel, following principles in \cite{strinati2024goal}, allows cross-layer coordination where the \gls{MEC} can instruct UE1 to downscale or maintain quality, based on instantaneous channel states, as depicted in Fig.~\ref{fig:software}. Each transmitted frame includes metadata indicating compression status, guiding the MEC to either apply \gls{GAI}-based upscaling or forward unchanged.
%\aca{
Within this framework, the \gls{GAI} module functions as a semantic service collocated with the \gls{MEC}. It enhances downscaled frames before forwarding to the receiving UE. For implementation convenience, this control channel is carried over the data plane but can be realized within the \textit{Application Plane} \cite{strinati2024goal}. The module is abstracted as a content-aware network function, integrated through the \gls{NEF}. 
Performing upscaling at the \gls{MEC} rather than UE2 offers significant advantages: GAI-based enhancement is computationally intensive, requiring resources unavailable on battery-powered UEs. Offloading this task to the \gls{MEC} node ensures efficient processing while preserving energy on the receiving device. Additionally, the \gls{DL} from the \gls{BS} to UE2 benefits from higher transmit power and beamforming capabilities of \gls{mMIMO}, enabling reliable delivery of high-bitrate upscaled streams. This architecture enables a proactive and cross-layer optimized approach compared to conventional \gls{PAA} strategies that adjust quality reactively \cite{kua2017survey}.

\subsection{GAI-Based Frame Processing in MEC}
To address the challenge of maintaining high video quality under latency constraints, the GAI block enhances downscaled frames that are transmitted when the bitrate cannot support full-resolution video. Under poor channel conditions (low SNR), the video frame $\mathbf{X}(t)$ is downscaled to $\mathbf{X}_{\text{down}}(t) \in \mathbb{R}^{m' \times n'}$, where $m' < m$ and $n' < n$. The downscaled frame is transmitted, and the MEC uses the GAI model $G(\cdot)$ to reconstruct the original frame $\mathbf{X}(t)$. The reconstructed frame $\hat{\mathbf{X}}(t)$ is obtained as:
\begin{equation}
\hat{\mathbf{X}}(t) = G(\mathbf{X}_{\text{down}}(t), \theta),
\end{equation}
where $\theta$ are the learnable parameters of the GAI model. The model is trained to minimize the reconstruction loss:
\begin{equation}
    \mathcal{L}_\text{GAI} = \mathbb{E} \left[ \| \mathbf{X}(t) - \hat{\mathbf{X}}(t) \|^2 \right].
\end{equation}
Deploying the GAI model at the MEC enables higher-quality video reconstruction from downscaled frames, reducing the required number of\glspl{PRB} while meeting latency constraints by offloading enhancement from the UE to the edge.

To assess the reconstructed video quality, we use two complementary metrics: PSNR and \gls{VMAF}.
The \gls{PSNR} is a widely used objective measure of similarity between the original and reconstructed frames, defined as:
\begin{equation}\label{eq:psnr}
\mathrm{PSNR} = 10 \cdot \log_{10} \left( \frac{\mathrm{MAX}_I^2}{\mathrm{MSE}} \right),
\end{equation}
where $\mathrm{MAX}_I$ is the maximum possible pixel value (e.g., 255 for 8-bit channels).  
The corresponding \gls{MSE} is given by:
\begin{equation}
\mathrm{MSE} = \frac{1}{mnC} \sum_{c=1}^{C} \sum_{i=1}^{m} \sum_{j=1}^{n} \left( x_{ij}^{(c)}(t) - \hat{x}_{ij}^{(c)}(t) \right)^2,
\end{equation}
where $C$ is the number of color channels, and $m \times n$ is the resolution of each frame.
\gls{VMAF}~\cite{7986143} provides a perceptual quality score by combining multiple features - such as detail preservation, structural similarity, and motion consistency - into a single value using a machine learning model trained on human-rated video datasets. The output ranges from 0 to 100, with higher scores indicating better perceptual quality. \gls{VMAF} is widely used in industry, including by Netflix, for evaluating user-perceived video quality.

\subsection{Modeling}
We consider two upscaling approaches: traditional methods (e.g. interpolation) $u_\text{trad}(\mathbf{X_{down}}(t))$ applied at the receiving UE and GAI-based $u_\text{GAI}(\mathbf{X_{down}}(t))$ where $u_\text{trad}$ applies a traditional method to upscale video frame, $\mathbf{X_{down}}(t)$, and is used on the receiving UE if no GAI is employed.
The alternative is to apply $u_\text{GAI}$ to upscale the video frame $\mathbf{X_{down}}(t)$ at the \gls{MEC}, resulting in an enhanced frame $\hat{\mathbf{X}}(t)$ which is forwarded to the receiving \gls{UE}.
It has been shown \cite{bui2024role} that applying \gls{GAI} improves the picture quality compared to traditional upscaling when a video frame has been compressed for transmission. We build on these findings and analyze the trade-off, particularly in terms of latency.
The latency model for the traditional approach is as follows: The traditional path transmits either $\mathbf{X}(t)$ or its downscaled version $\mathbf{X}_{\text{down}}(t)$, followed by UE-side upscaling:
\begin{equation} \label{eq:latmodel}
L_\text{trad} = t_\text{RAN}(\mathbf{X}_\text{tx}(t)) 
+ t_\text{CN}(\mathbf{X}_\text{tx}(t)) 
+ t_\text{UE2}(\mathbf{X}_\text{tx}(t)),
\end{equation}
where $\mathbf{X}_\text{tx}(t)\in\{\mathbf{X}(t),\mathbf{X}_{\text{down}}(t)\}$ 
denotes the time to transmit the frame over the RAN \gls{UL} channel, $t_\text{CN}(\mathbf{X}(t))$ reflects the time required to pass through the \gls{CN}, and $t_\text{UE2}(\mathbf{X}(t))$ denotes the \gls{DL} latency for the receiving UE, including further processing. This processing includes upscaling of the video frame if it has been compressed.

Our GAI-based approach includes additional MEC processing and enhanced frame transmission: 
\begin{equation}\label{eq}
\begin{split}
L_\text{GAI} = t_\text{RAN}(\mathbf{X}_{tx}(t)) + t_\text{CN}(\mathbf{X}_{tx}(t)) + \\ t_\text{MEC}(\mathbf{X'}_{tx}(t)) 
+ t_\text{CN}(\mathbf{X'}_{tx}(t)) + t_\text{UE2}(\mathbf{X'}_{tx}(t)).
\end{split}
\end{equation}
where $\mathbf{X'}_\text{tx}(t)\in\{\mathbf{X}(t),\hat{\mathbf{X}}(t)\}$; specifically, $\mathbf{X'}_\text{tx}(t)=\hat{\mathbf{X}}(t)$ only when a downscaled frame $\mathbf{X}_{\text{down}}(t)$ is transmitted. In practice,  this occurs under poor channel conditions (low SNR and limited bitrate), whereas otherwise the MEC simply forwards $\mathbf{X}(t)$ without additional 
processing.

\begin{figure}[t!]
	\centering
	\includegraphics[width=0.45\textwidth]{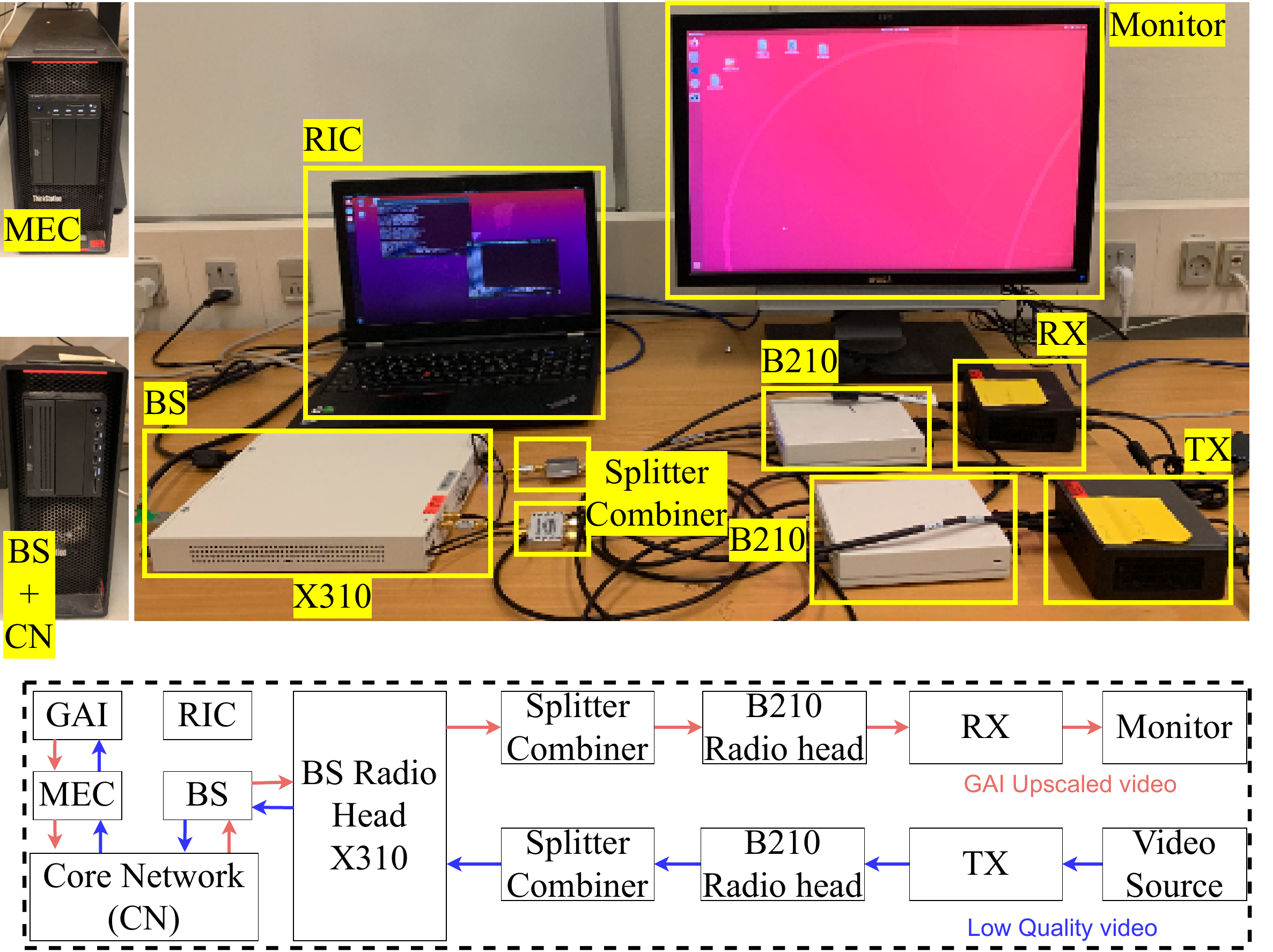}
	\caption{Experimental lab setup.}
	\label{fig:lab-setup}
  \vspace{-15pt}
\end{figure}

\section{Experimental Setup}\label{sec:results}
We extend the ORAN experimental platform in~\cite{bui2024role} to evaluate our proposed methodology. We analyze the impact of our \gls{GAI}-upscaling solution on \gls{E2E} video latency and its behavior in a live testbed setting.
For video enhancement, we employ RealSR \cite{ji2020real} as our \gls{GAI} model. RealSR was selected due to its balance between inference speed and perceptual quality, which makes it feasible to deploy on our edge hardware for real-time computing. 
The platform shown in Fig.~\ref{fig:lab-setup} uses OpenAirInterface to orchestrate the RAN, with an eNB at the base station. For the \glspl{UE} srsRAN is used, Open5GS handles core network functions, and the RIC is implemented using FlexRIC. 
A computer equipped with an NVIDIA RTX 4080 GPU serves as the MEC server running the GAI model.
Clocks on all nodes are synchronized with Chrony for sub-ms latency measurements.

In cases where bad channel quality leads to UE1 transmitting downscaled video frames, the \gls{GAI} module is employed to enhance the video quality before forwarding to the receiving \gls{UE} as seen in Fig.~\ref{fig:lab-setup}. Otherwise, the video feed is forwarded without modifications.
To trigger compression activation, we vary TX power periodically, switching every 15 seconds. This produces alternating high and low \gls{UL} \gls{MCS} periods, resulting in corresponding bitrate fluctuations. The system, unaware of this switching schedule, responds based on observed channel status reported by the xApp.
%\aca{
This periodic switching serves as a controlled and repeatable test rather than a detailed channel model. It emulates transitions in RAN conditions that trigger application adaptation, allowing consistent comparison between the approaches. The controller relies solely on RAN feedback. While strictly reactive to instantaneous metrics, it operates proactively relative to conventional feedback-based methods by acting on RAN information before congestion manifests at the transport or application layer.%}
To measure \gls{E2E}  latency, we developed two Python programs for video transmission and reception, deployed on UE1 and UE2, respectively, to represent a video streaming scenario.
The transmission program on UE1 integrates the semantic control channel (described in Section~\ref{sec:system}), enabling the \gls{MEC} to control compression methods and video resolution. This program logs timestamps when video frames become available before size reduction.
The reception program on UE2 receives video frames and logs timestamps after resizing frames via traditional upscaling, when applicable.
Additionally, the \gls{GAI}-module continuously logs channel \gls{SNR} every 1 millisecond through the xApp. These logs serve two purposes: real-time control decisions for instructing UE1, and offline performance analysis of our system.

\begin{figure}[t!]
	\centering
    \includegraphics[width=0.45\textwidth]{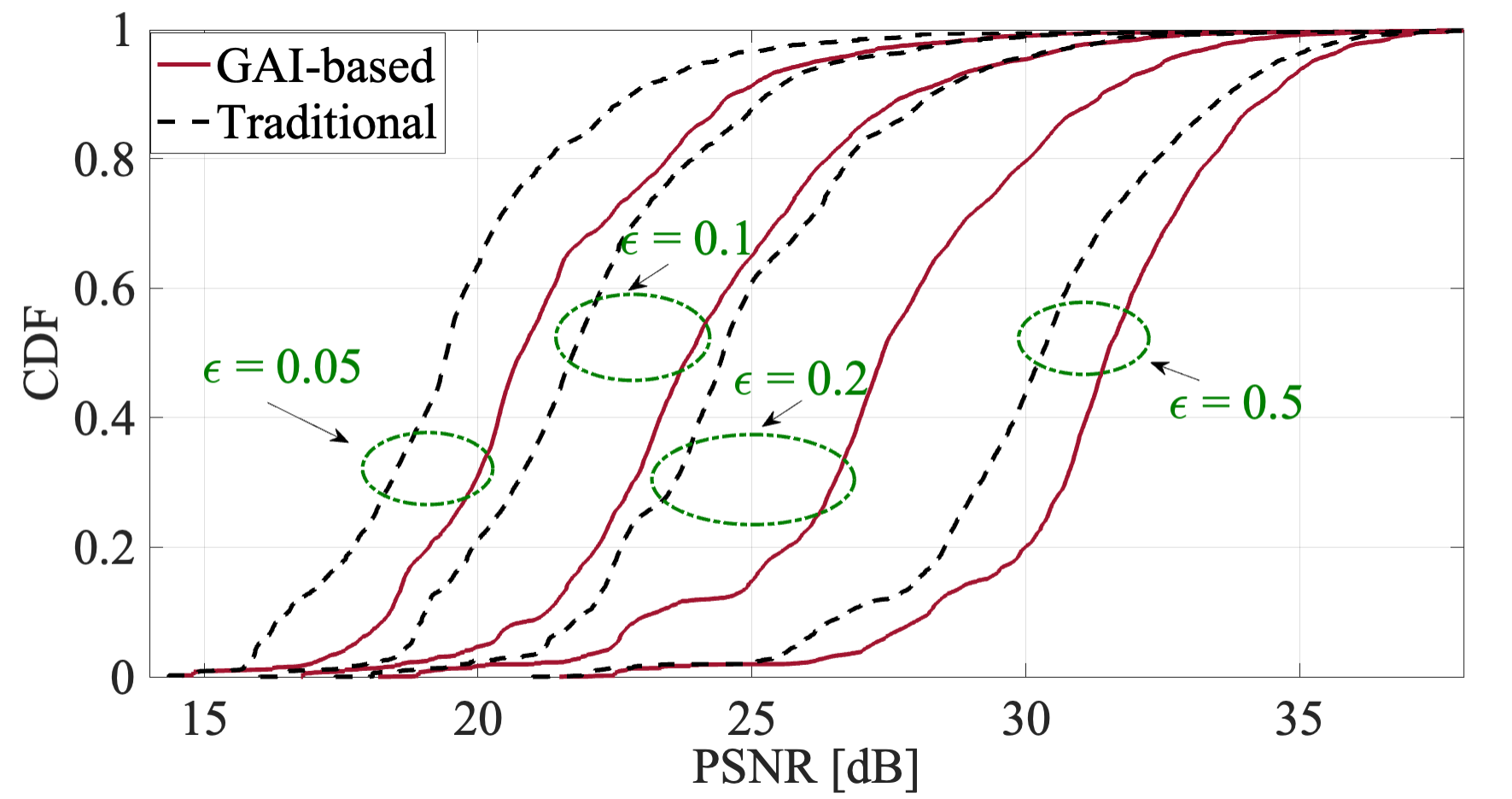}
	\caption{CDF of PSNR under different downscaling factors ($\epsilon$).}
	\label{fig:oran-experiment}
	\vspace{-15pt}
\end{figure}

\section{Results and Discussion}
\label{sec:Numerical}
Building on the system model and architecture, this section demonstrates the effectiveness of the GAI-based approach in improving video quality and managing latency under varying channel conditions. Evaluation is conducted on a live ORAN testbed, assessing both perceptual and network-level metrics. 
%\aca{
The experiments explore how compute resources for GAI-based upscaling can compensate for reduced communication capacity. Perceptual fidelity is analyzed across multiple compression levels, where $\epsilon$ denotes the fraction of retained pixels (e.g., $\epsilon$ = 0.2 keeps 20\% of the original pixels), while latency is evaluated for an exemplary compression setting to compare GAI and traditional upscaling.%}
We compare our proposed congestion control using an xApp with GAI upscaling (\textit{PCC-GAI}) against a conventional \gls{PAA} baseline. In \textit{PCC-GAI}, the receiver monitors early signs of link degradation and proactively signals a bitrate reduction, while monitoring the capacity to revert to the original quality. The reduced stream is then reconstructed at the receiver using GAI-based upscaling to restore perceptual detail. In contrast, \gls{PAA} adjusts the bitrate only after congestion becomes apparent and probes for recovery. It relies on traditional upscaling, with comparatively reduced visual fidelity. This means \textit{PCC-GAI} responds faster to capacity variations and maintains visual quality under constrained bitrate, whereas \gls{PAA} reacts more slowly and produces visibly lower-quality frames under the same conditions.

Fig.~\ref{fig:oran-experiment} illustrates the CDF of \gls{PSNR} between received and transmitted frames by deploying our strategy (\textit{PCC-GAI}) versus the \gls{PAA} scheme (normal transmit and receive frames) under different downscaling factors, $\epsilon$. Specifically, with $\epsilon = 0.2$, there are 60\% frames of \textit{Traditional} scaling with a PSNR below $25$~dB. Thanks to the advanced \textit{GAI-based}, the proposed architecture provides less than 20\% of the frames having a PSNR below $25$~dB. This demonstrates that integrating GAI-based reconstruction into the adaptation loop significantly improves perceptual quality at reduced bitrates.

\begin{figure}[t]
	\centering
    \includegraphics[width=0.45\textwidth]{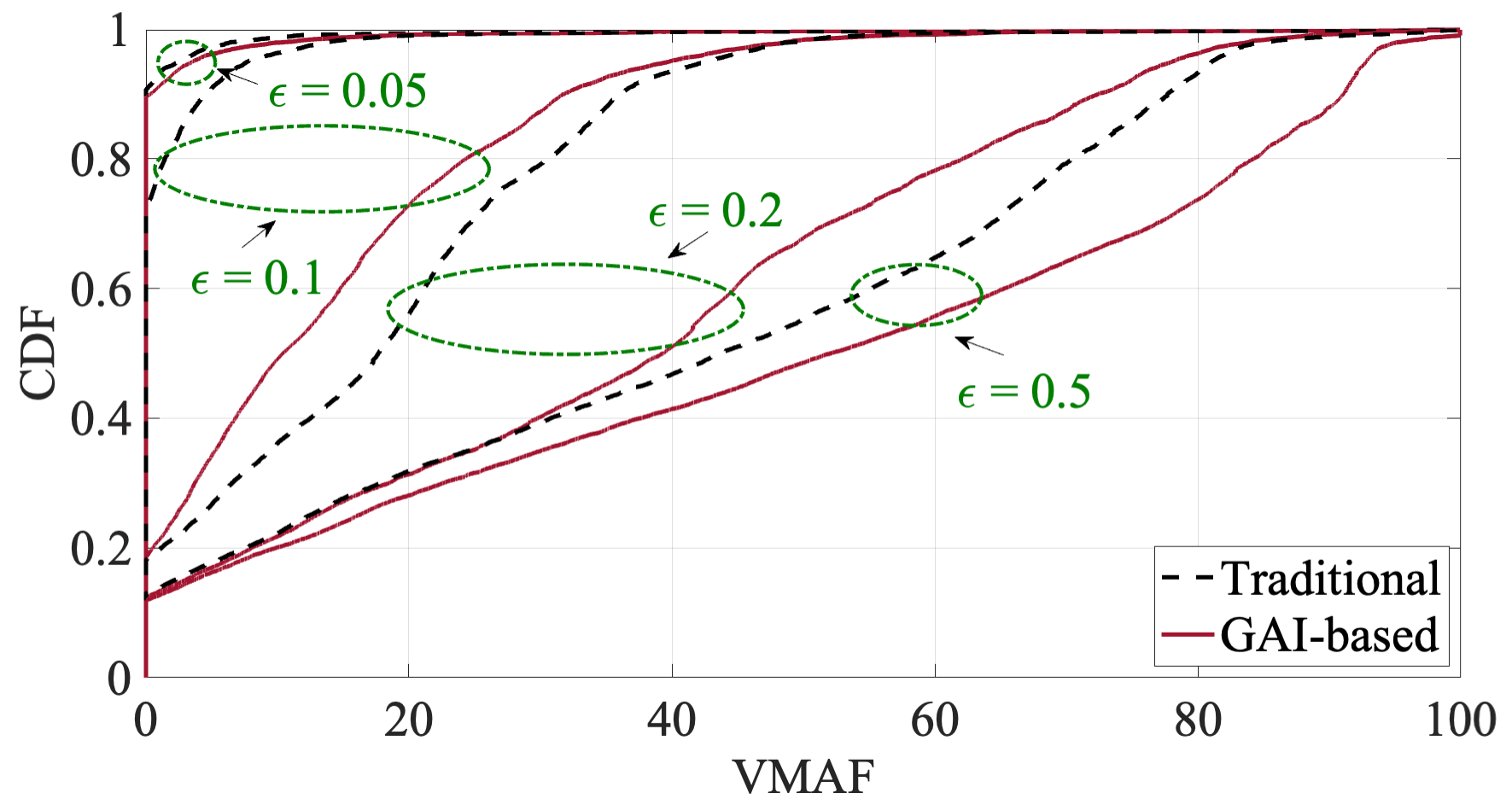}
	\caption{CDF of VMAF for different downscaling factors ($\epsilon$).}
	\label{fig:CDF_VMAF}
	 \vspace{-15pt}
\end{figure}

Fig.~\ref{fig:CDF_VMAF} illustrates the CDF of the \gls{VMAF} metric across $\epsilon$, comparing the performance of the \textit{GAI-based} method with \textit{Traditional} scaling. As observed, under tight resolution constraints (e.g., $\epsilon = 0.05\text{--}0.2$), the \textit{GAI-based} method significantly outperforms the traditional baseline, maintaining higher \gls{VMAF} values over a larger fraction of frames. For instance, at $\epsilon = 0.1$, approximately 100\% of frames processed using the traditional pipeline yield \gls{VMAF} scores below 20, while the \textit{GAI-based} system maintains considerably higher \gls{VMAF} distributions, suggesting perceptually improved quality even under aggressive compression. As $\epsilon$ increases to 0.2 and 0.5, indicating milder compression, both approaches improve in perceptual quality; still, the advantage of \gls{GAI} persists. This demonstrates the resilience of GAI-based reconstruction under both extreme and moderate bitrate constraints.

To evaluate latency performance experimentally, we transmitted 50 randomly selected videos from \cite{ahmed2023malicious} through our testbed. Fig.~\ref{fig:latency} compares two congestion control: \textit{PCC-GAI} (with GAI upscaling) and \gls{PAA} (with traditional upscaling). For the latency-specific comparison, we also illustrate \textit{PCC-noGAI-based}, which uses \gls{PCC} for congestion control and traditional upscaling.
In \textit{PCC-GAI} and \textit{PCC-noGAI}, the \gls{MEC} controller proactively instructs the transmitting UE to downscale resolution based on \gls{ORAN} control feedback. This resolution adaptation introduces a delay, as GAI processing is more computationally demanding than simply forwarding video frames. In contrast, the \gls{PAA} method transmits frames immediately and adjusts the bitrate reactively in response to observed congestion.
In Fig.~\ref{fig:latency}, the \gls{PAA} has a larger fraction of low latencies due to low-resolution transmission without \gls{GAI} processing. However, after approximately 100~ms (around the 80th percentile of the CDF), it exhibits latency spikes, with some frames exceeding 600~ms. This is a result of buffering caused by the reactive behavior of the \gls{PAA}-method, which also leads to a larger fraction of high latencies.
Examining the control channel behavior, we see that \textit{PCC-noGAI} has a similar latency profile to the \gls{PAA} for low latencies, and at higher latencies becomes similar to \textit{PCC-GAI}, yielding the best of both approaches in terms of latency (however, without the visual improvements of GAI upscaling).
The latency degradation for the \gls{PAA} is attributed to buffer congestion caused by delayed adaptation when the signal quality is reduced. In contrast, \textit{PCC-GAI}, despite its initial overhead, avoids more late-stage delays by proactively reducing video resolution, resulting in lower tail latencies and more stable performance overall.
In conjunction with the results in Fig.~\ref{fig:CDF_VMAF}, these findings show that proactive, compute-heavy strategies like GAI improve both latency stability and video quality. Although this comes at a processing cost, the dual benefit of reduced delay variance and enhanced visual fidelity supports its use in latency- and video quality-sensitive applications.
\begin{figure}[t]
    \centering
    \input{figs/tikz/latency_cdf_new}
    \vspace{-15pt}
\caption{CDF of \gls{E2E} latency for the different methods (\(\varepsilon = 0.0625\)).}
    \label{fig:latency}
    \vspace{-15pt}
\end{figure}
Fig.~\ref{fig:time-view} presents a time-series view of channel SNR and the resulting video frame PSNR for the two scaling methods, taken from a segment of the video. The data comes from two independent experiment runs performed at the same video timestamp, with identical target changes in channel \gls{SNR} (a drop and subsequent recovery), ensuring the runs are time-aligned. While the instantaneous SNR may vary slightly in each run due to random factors, the overall patterns are statistically similar, enabling a meaningful comparison.
In the first and last segments - where channel \gls{SNR} is relatively low - the \gls{PSNR} achieved by the \textit{GAI-based} method is higher than that of the \gls{PAA} scaling, indicating better image quality. Notably, during some low-SNR intervals, the traditional approach exhibits slightly higher instantaneous channel \gls{SNR}, yet still results in lower \gls{PSNR}, further underscoring the benefit of \gls{GAI}-based scaling.

This analysis shows that GAI integration improves video quality under constrained conditions, while \gls{PCC} reduces \gls{E2E}  latency despite computational overhead. The combined benefits make the approach suitable for dynamic environments such as dense urban deployments, where MEC offloading can reduce \gls{UL} RAN load while preserving user-perceived quality.
However, the approach increases MEC inference load. Our testbed supported real-time execution, but practical deployments may share MEC resources across multiple latency-sensitive applications, potentially introducing delays.
Codec selection was abstracted to isolate the GAI upscaling effect and ensure a fair \gls{PAA} comparison; thus, the reported latency results are codec-agnostic. In practice, compression introduces trade-offs between encoding/decoding delay and reduced transmission time, with the net effect depending on codec efficiency and MEC resource availability.
A complete characterization of the bitrate–compute–quality tradeoff could be formalized as an optimization problem under resource constraints. We note this as a direction for future work, while the present study focuses on experimentally demonstrating the networking functionality and its computational tradeoffs.

\begin{figure}
    \centering
    \input{figs/tikz/SNR_timeview_old}
    \vspace{-5pt}
    \caption{Time-series view of channel SNR and resulting PSNR for GAI-based and Traditional upscaling (\(\varepsilon = 0.05\)).}
    \label{fig:time-view}
     \vspace{-15pt}
\end{figure}

\section{Conclusion}\label{sec:conclusion}
\glsresetall
This work demonstrated two key contributions for enhancing live video streaming through \gls{ORAN}. First, we introduced a semantic control channel between the \gls{MEC} and \gls{UE} that enables proactive adaptation to channel dynamics. Unlike traditional application-level reactive adaptation, this control channel reduces buffering and \gls{E2E}  latency for both \gls{GAI}-based and \gls{PAA}-based streaming by triggering adaptation before congestion manifests at the application layer. Second, we integrated \gls{GAI}-based upscaling at the \gls{MEC} to maintain high perceived video quality under \gls{UL}-limited conditions, achieving up to 4 dB PSNR and 15 points VMAF improvement while eliminating latency spikes exceeding 600 ms.
Our results highlight the value of cross-layer coordination across the \gls{RAN}, application, and \gls{MEC} layers, demonstrating how \gls{ORAN}-enabled control improves both network efficiency and end-user experience. This approach offers a viable path for future latency- and video-quality-sensitive applications. While our evaluation used a pre-trained RealSR model, the performance of any \gls{GAI}-based approach depends on the representativeness of its training data, and domain mismatch remains a challenge for practical deployment.

\setstretch{0.9}
%\bibliographystyle{IEEEtran}
% \vspace*{-10pt}
%\bibliography{Journal}
% Generated by IEEEtran.bst, version: 1.12 (2007/01/11)

 % For arxiv

\end{document}